\documentclass[fleqn,useAMS,usenatbib]{mn2e}
\usepackage{graphicx}
\usepackage{amsmath}
\usepackage{amssymb}
\title[Variance of a 3D physical field from 2D observations]{A method for reconstructing the variance of a 3D physical field from 2D observations: Application to turbulence in the ISM}
\author[C. M. Brunt, C. Federrath, \& Price, D. J.]{C. M. Brunt$^{1}$\thanks{E-mail brunt@astro.ex.ac.uk}, C. Federrath$^{2,3}$, \& D. J. Price$^{4}$\\
$^{1}$School of Physics, University of Exeter, Stocker Road, Exeter, UK\\
$^{2}$Zentrum f\"{u}r Astronomie der Universit\"{a}t Heidelberg, Institut f\"{u}r Theoretische Astrophysik, Albert-Ueberle-Str. 2, \\ D-69120 Heidelberg, Germany\\
$^{3}$Max-Planck-Institute for Astronomy, K\"{o}nigstuhl 17, D-69117 Heidelberg, Germany \\
$^{4}$Centre for Stellar and Planetary Astrophysics, School of Mathematical Sciences, Monash University, Clayton Vic 3168, Australia}
\begin{document}

\date{Accepted ; Received ; in original form }

\pagerange{\pageref{firstpage}--\pageref{lastpage}} \pubyear{2009}

\maketitle

\label{firstpage}

\begin{abstract}
We introduce and test an expression for calculating the variance of a physical field in three dimensions
using only information contained in the two-dimensional projection of the field. The method is general
but assumes statistical isotropy. To test the method we apply it to numerical simulations of 
hydrodynamic and magnetohydrodynamic turbulence in molecular clouds, and demonstrate that it can
recover the 3D normalised density variance with $\sim$~10\% accuracy if the assumption of isotropy is valid.
We show that the assumption of isotropy breaks down at low sonic Mach number if the turbulence is
sub-Alfv\'{e}nic. Theoretical predictions suggest that the 3D density variance
should increase proportionally to the square of the Mach number of the turbulence. 
Application of our method will allow this prediction
to be tested observationally and therefore constrain a large body of analytic models of star formation 
that rely on it. 
\end{abstract}


\section{Introduction}

It is a fundamental problem in astrophysics that we typically only have access to two-dimensional (2D) physical fields that have been
integrated or averaged over the line-of-sight, while the physical fields of interest are intrinsically
three-dimensional (3D). Constraining 3D properties is very difficult, and is limited to structures with a fairly simple
geometry (e.g. Lucy 1974 -- see Reblinsky 2000 for an application to galaxy clusters). Molecular clouds have
very complex structure that is not suitable for direct inversion from 2D to 3D, but there exist methods for inferring statistical
information on the scaling behaviour of the 3D structure of the density and velocity fields (e.g. Stutzki et al 1998). Molecular 
cloud evolution is driven in large part by the action of turbulence, and recently, much interest has been directed towards the action 
of turbulence in shaping the density field in molecular clouds, in particular towards the theoretically predicted increase in the variance 
of the density field with the Mach number of the turbulence (Padoan, Nordlund, \& Jones 1997; Passot \& V\'{a}zquez-Semadeni 1998).
This prediction plays a key role in analytic models of star formation (Padoan \& Nordlund 2002; Krumholz \& McKee 2005; Elmegreen 2008; 
Hennebelle \& Chabrier 2008; Padoan \& Nordlund 2009; Hennebelle \& Chabrier 2009). 

If the density, $\rho$, is expressed in units of the mean density, $\rho_{0}$, the theoretical predictions for the 
relationship between the density variance and root mean square (rms) Mach number, $M$, may be simply written as:
\begin{equation}
\sigma_{\rho / \rho_{0}}^{2} \; = \; b^{2}M^{2} ,
\label{eqno0}
\end{equation}
where $b$ is a constant of proportionality.
In the case of isothermal gas, which is applicable to a good approximation to molecular clouds, the probability density
function (PDF) of the density field is thought to be lognormal in form (V\'{a}zquez-Semadeni 1994). Using a lognormal form 
for the density PDF, Padoan \& Nordlund (2002) derive a relation between the variance in the logarithm of the 
density field, $\sigma_{\ln{(\rho / \rho_{0}})}^{2}$, and the rms Mach number, $M$:
\begin{equation}
   \sigma_{\ln{(\rho / \rho_{0}})}^{2} =\ln{(1 + b^{2}M^{2})} ,
\label{eqno1}
\end{equation}
which is equivalent to equation~(\ref{eqno0}). A range of predictions for $b$ have been proposed,
which were recently synthesized into a unified model by Federrath et al (2008) who propose that $b = 1/3$ 
for solenoidal (divergence-free) forcing and $b = 1$ for compressive (curl-free) forcing in 3D.

Given the importance of equation~(\ref{eqno0}) for analytic models of star formation, it is essential to test it 
with observational data. Initial observational tests by Goodman, Pineda, \& Schnee (2009) did not find any 
obvious support for the theoretical predictions. Federrath et al (2009) have suggested a reason for this
lack of agreement, citing variations in $b$ caused by different turbulent forcing mechanisms. 
Perhaps more importantly, these observations relied on measuring the variance in the projected 2D column density rather 
than in the 3D density field that appears in the theoretical predictions. From observations, we do not have access to the 
3D density and velocity fields to directly test this prediction, and evaluate its applicability to molecular 
clouds and to the predictive star formation models. 

In this paper, we derive and test an expression that relates the observable variance in the 2D column density field to the true
3D variance of the density field. The method is completely general, albeit with the assumption of statistical
isotropy, and may also be applied to the projection of other physical fields. In Section 2, we present the analytical
expressions needed to convert the 2D variance into the 3D variance. In Section 3, we demonstrate the
application of the method to numerical simulations of hydrodynamic and magnetohydrodynamic turbulence. Our summary is given
in Section 4. 

\section[]{The Statistics of Projection from 3D to 2D}

\subsection{Development of the Method}

Given a physical field in 3D, we wish to examine how the statistical properties of a 2D 
projected field are related to the intrinsic properties. By doing this, we can then infer 
intrinsic quantities from the measureable quantities. We focus here on the variance,
measured in 3D and in projected 2D. In our initial derivation, we consider a 3D field, $F_{3}(x,y,z)$,
which is {\it averaged} along the $z$ axis to produce a projected field $F_{2}(x,y)$, via:
\begin{equation}
 F_{2}(x,y) = {\frac{1}{L}} \int_{-L/2}^{L/2} dz \;  F_{3}(x,y,z)
\label{eqno2}
\end{equation}
where $L$ is the physical size of the region, assumed cubical, that contains $F_{3}$.

We will make use of the 3D Fourier series of $F_{3}$, obtained over the interval $[-L/2,+L/2]$, which is:
\begin{equation}
\tilde{F_{3}}({\mathbfit{k}}) =  \displaystyle\int_{-L/2}^{L/2}\displaystyle\int_{-L/2}^{L/2}\displaystyle\int_{-L/2}^{L/2} \; d^{3}\mathbfit{\mathbfit{r}} \; F_{3}({\mathbfit{r}}) \; e^{ -2 \pi i {\mathbfit{k}} \cdot  {\mathbfit{r}} / L}
\label{eqno3}
\end{equation}
where $\mathbfit{\mathbfit{r}} = (x\hat{\mathbfit{x}},y\hat{\mathbfit{y}},z\hat{\mathbfit{z}})$ and 
the spatial frequencies are 
$\mathbfit{k} = (k_{x}\hat{\mathbfit{x}},k_{y}\hat{\mathbfit{y}},k_{z}\hat{\mathbfit{z}})$, for integer
$k_{x}$, $k_{y}$, and $k_{z}$.
The field $F_{3}$ can then be written as:
\begin{equation}
{F_{3}}({\mathbfit{r}}) = {\frac{1}{L^{3}}} \displaystyle\sum_{k_{x}=-\infty}^{\infty} \sum_{k_{y}=-\infty}^{\infty} \sum_{k_{z}=-\infty}^{\infty} \; \tilde{F_{3}}(k_{x},k_{y},k_{z}) \; e^{ 2 \pi i {\mathbfit{k}} \cdot  {\mathbfit{r}} / L} .
\label{eqno4}
\end{equation}
Inserting equation~(\ref{eqno4}) into equation~(\ref{eqno2}),
\begin{eqnarray}
\lefteqn{ F_{2}(x,y) = } \nonumber \\
\lefteqn{ {\frac{1}{L^{4}}} \displaystyle\int_{-L/2}^{L/2} dz \displaystyle\sum_{k_{x}=-\infty}^{\infty} \sum_{k_{y}=-\infty}^{\infty} \sum_{k_{z}=-\infty}^{\infty} \tilde{F_{3}}(k_{x},k_{y},k_{z}) \; e^{ 2 \pi i {\mathbfit{k}} \cdot  {\mathbfit{r}} / L} }
\label{eqno5}
\end{eqnarray}
and computing the $z$ integral first, we find that:\begin{equation}
F_{2}(x,y) = {\frac{1}{L^{3}}} \displaystyle\sum_{k_{x}=-\infty}^{\infty} \sum_{k_{y}=-\infty}^{\infty} \tilde{F_{3}}(k_{x},k_{y},k_{z}=0) \; e^{ 2 \pi i {\mathbfit{k}} \cdot  {\mathbfit{r}} / L} ,
\label{eqno6}
\end{equation}
since
\begin{equation}
{\frac{1}{L}} \int_{-L/2}^{L/2} dz \; e^{2 \pi i k_{z} z / L} = 1 \;\; {\mathrm{for}} \;\; k_{z} = 0 , 
\end{equation}
\begin{equation}
{\frac{1}{L}} \int_{-L/2}^{L/2} dz \; e^{2 \pi i k_{z} z / L} = 0 \;\; {\mathrm{for}} \;\; k_{z} \neq 0 .
\end{equation}
Since the 2D Fourier series, over the interval $[-L/2,+L/2]$, of $F_{2}$ is:
\begin{equation}
\tilde{F_{2}}({\mathbfit{k$_{2}$}}) =  \displaystyle\int_{-L/2}^{L/2}\displaystyle\int_{-L/2}^{L/2} \; d^{2}{\mathbfit{r$_{2}$}} \; F_{2}({\mathbfit{r$_{2}$}}) \; e^{ -2 \pi i {\mathbfit{k$_{2}$}} \cdot  {\mathbfit{r$_{2}$}} / L}
\label{eqno9}
\end{equation}
where $\mathbfit{\mathbfit{r$_{2}$}} = (x\hat{\mathbfit{x}},y\hat{\mathbfit{y}})$ and
$\mathbfit{k$_{2}$} = (k_{x}\hat{\mathbfit{x}},k_{y}\hat{\mathbfit{y}})$, and:
\begin{equation}
{F_{2}}({\mathbfit{r$_{2}$}}) = {\frac{1}{L^{2}}} \displaystyle\sum_{k_{x}=-\infty}^{\infty} \sum_{k_{y}=-\infty}^{\infty} \tilde{F_{2}}(k_{x},k_{y}) \; e^{ 2 \pi i {\mathbfit{k$_{2}$}} \cdot  {\mathbfit{r$_{2}$}} / L} ,
\label{eqno10}
\end{equation}
we see that, comparing equation~(\ref{eqno6}) and equation~(\ref{eqno10}):
\begin{equation}
\tilde{F_{2}}(k_{x},k_{y}) = {\frac{1}{L}} \tilde{F_{3}}(k_{x},k_{y},k_{z}=0) .
\label{eqno11}
\end{equation}
 
In other words: the 2D Fourier series of $F_{2}$ is proportional to the $k_{z} = 0$ cut through $\tilde{F_{3}}$.
Previous studies have made use of this result (e.g. Stutzki et al. 1998; Lazarian \& Pogosyan 2000; Brunt \& Mac Low 2004;
Miville-Deschenes \& Martin (2007)). As the power spectrum, $P_{3}({\mathbfit{k}})$, is the squared modulus of the
Fourier transform ($P_{3}({\mathbfit{k}}) = \tilde{F_{3}}({\mathbfit{k}}) \tilde{F_{3}}^{*}({\mathbfit{k}})$), 
2D power spectra of projected fields have been used to infer the 3D power 
spectrum, under the assumption of isotropy, i.e. that the $k_{z} = 0$ cut through the power spectrum
is statistically representive of the full power spectrum. In our analysis below, we will also make use of the power spectrum
and the assumption of isotropy. 

Now we compute the variances in the 3D and 2D fields. From the mean value of $F_{3}$:
\begin{equation}
\langle {F_{3}} \rangle = {\frac{1}{L^{3}}} \displaystyle\int_{-L/2}^{L/2}\displaystyle\int_{-L/2}^{L/2}\displaystyle\int_{-L/2}^{L/2} \; d^{3}{\mathbfit{r}} \; F_{3}({\mathbfit{r}})  ,
\end{equation}
and the mean square value of $F_{3}$:
\begin{equation}
\langle {F_{3}^{2}} \rangle = {\frac{1}{L^{3}}} \displaystyle\int_{-L/2}^{L/2}\displaystyle\int_{-L/2}^{L/2}\displaystyle\int_{-L/2}^{L/2} \; d^{3}{\mathbfit{r}} \; F_{3}^{2}({\mathbfit{r}}) ,
\end{equation}
we find the variance of $F_{3}$ as:
\begin{equation}
\sigma_{3}^{2} = \langle {F_{3}^{2}} \rangle - \langle {F_{3}} \rangle^{2} .
\label{eqno12}
\end{equation}
We also make use of the Fourier transform of $F_{3}$, by noting that:
\begin{equation}
\langle {F_{3}} \rangle = {\frac{1}{L^{3}}} \tilde{F_{3}}(0,0,0) 
\end{equation}
and, through Parseval's Theorem:
\begin{equation}
\langle {F_{3}^{2}} \rangle = {\frac{1}{L^{6}}} \displaystyle\sum_{k_{x}=-\infty}^{\infty} \sum_{k_{y}=-\infty}^{\infty} \sum_{k_{z}=-\infty}^{\infty} \; \tilde{F_{3}}\tilde{F_{3}}^{*}  ,
\end{equation}
allowing us to rewrite equation~(\ref{eqno12}) as:
\begin{equation}
\sigma_{3}^{2} = {\frac{1}{L^{6}}} \left( \left( \displaystyle\sum_{k_{x}=-\infty}^{\infty} \sum_{k_{y}=-\infty}^{\infty} \sum_{k_{z}=-\infty}^{\infty} \; \tilde{F_{3}}\tilde{F_{3}}^{*} \right) - \tilde{F_{3}}^{2}(0,0,0) \right) .
\label{eqno17}
\end{equation}
By similar analysis, we find the variance of $F_{2}$ is:
\begin{equation}
\sigma_{2}^{2} = {\frac{1}{L^{4}}} \left( \left( \displaystyle\sum_{k_{x}=-\infty}^{\infty} \sum_{k_{y}=-\infty}^{\infty} \; \tilde{F_{2}}\tilde{F_{2}}^{*} \right)  - \tilde{F_{2}}^{2}(0,0) \right),
\end{equation}
or, applying equation~(\ref{eqno11}):
\begin{eqnarray}
\lefteqn{ \sigma_{2}^{2} = {\frac{1}{L^{6}}} \times } \nonumber \\
\lefteqn{ \left( \left( \displaystyle\sum_{k_{x}=-\infty}^{\infty} \sum_{k_{y}=-\infty}^{\infty} \; \tilde{F_{3}}(k_{z}=0)\tilde{F_{3}}^{*}(k_{z}=0) \right) - \tilde{F_{3}}^{2}(0,0,0) \right). } \nonumber \\
\label{eqno19}
\end{eqnarray}
The ratio, $R$, of the variance of $F_{2}$ to the variance of $F_{3}$ is therefore:
\begin{eqnarray}
\lefteqn{ \frac{\sigma_{2}^{2}}{\sigma_{3}^{2}} = R = } \nonumber \\
\lefteqn{ \frac{ \left( \displaystyle\sum_{k_{x}=-\infty}^{\infty} \sum_{k_{y}=-\infty}^{\infty} \; \tilde{F_{3}}(k_{z}=0)\tilde{F_{3}}^{*}(k_{z}=0) \right) - \tilde{F_{3}}^{2}(0,0,0)  }{ \left( \displaystyle\sum_{k_{x}=-\infty}^{\infty} \sum_{k_{y}=-\infty}^{\infty} \sum_{k_{z}=-\infty}^{\infty} \; \tilde{F_{3}}\tilde{F_{3}}^{*} \right)  - \tilde{F_{3}}^{2}(0,0,0) } } . \nonumber \\
\label{eqno20}
\end{eqnarray}

In practice, the measureable physical fields, either through observations or numerical
simulations, will consist of a discrete number of measurements at a fixed set of grid
points. We define the scale ratio, $\lambda$, as the ratio of the image (or cube) size
to the pixel size. Taking the grid to be of size 
$\lambda \times \lambda \times \lambda$ or $\lambda \times \lambda$ pixels 
in 3D and 2D respectively, the Fourier transforms are carried out at a discrete set
of spatial frequencies, $k = -\lambda/2 + 1, -\lambda/2+2, ... , -2, -1, 0, 1, 2 ... , \lambda/2-1, \lambda/2$ along each axis.
The quantity $\tilde{F_{3}}\tilde{F_{3}}^{*}$ is the spectral power, $P_{3}$, which can
be observationally derived (up to an unimportant constant of proportionality) as follows. If we had a projected
mean 2D field, $F_{2}$, and wanted to use this to infer the 3D variance, then we would compute its power spectrum, 
$P_{2}(k_{x},k_{y})$, and from this produce an azimuthally-averaged power spectrum $P_{2}(k)$, which depends only
on the modulus of the spatial frequency, $k$. The key idea behind the method is that we can, assuming isotropy, take:
\begin{equation}
P_{3}(k) \propto P_{2}(k) , 
\end{equation}
to obtain:
\begin{equation}
R = \frac{ \left( \displaystyle\sum_{k_{x}=-\lambda/2+1}^{\lambda/2} \sum_{k_{y}=-\lambda/2+1}^{\lambda/2} P_{2}(k) \right) - P_{2}(0)}{ \left( \displaystyle\sum_{k_{x}=-\lambda/2+1}^{\lambda/2} \sum_{k_{y}=-\lambda/2+1}^{\lambda/2} \sum_{k_{z}=-\lambda/2+1}^{\lambda/2} P_{2}(k) \right) - P_{2}(0)} ,
\label{eqno21}
\end{equation}
The 3D variance can then be calculated as $\sigma_{3}^{2} = \sigma_{2}^{2}/R$. Note however that since $\lambda$ is 
necessarily finite, the observed 2D variance and the calculated 3D variance are lower limits to the true variances 
that would be obtained in the limit $\lambda \longrightarrow \infty$. This is discussed further below.

A more compact form of equation~(\ref{eqno21}) can be obtained by defining:
\begin{equation}
{\displaystyle\sum_{k \neq 0}}^{2D,\lambda} P_{2}(k) = \left( \displaystyle\sum_{k_{x}=-\lambda/2+1}^{\lambda/2} \sum_{k_{y}=-\lambda/2+1}^{\lambda/2} P_{2}(k) \right) - P_{2}(0)
\label{eqno21aa}
\end{equation}
and:
\begin{eqnarray}
\lefteqn{ {\displaystyle\sum_{k \neq 0}}^{3D,\lambda} P_{2}(k) = } \nonumber \\
\lefteqn{ \left( \displaystyle\sum_{k_{x}=-\lambda/2+1}^{\lambda/2} \sum_{k_{y}=-\lambda/2+1}^{\lambda/2}\sum_{k_{z}=-\lambda/2+1}^{\lambda/2}  P_{2}(k) \right) - P_{2}(0) } \nonumber \\
\label{eqno21bb}
\end{eqnarray}
so that:
\begin{equation}
R = \frac{{\displaystyle\sum_{k \neq 0}}^{2D,\lambda} P_{2}(k)}{{\displaystyle\sum_{k \neq 0}}^{3D,\lambda} P_{2}(k)}
\label{eqno21cc}
\end{equation}

For this method to work, it is essential that the projected field, $F_{2}$, is the line-of-sight
{\it average} of the 3D field, $F_{3}$ (see equation~(\ref{eqno2})). In many instances, the projected 2D field is the line-of-sight
{\it intregral} of the 3D field (e.g. column density versus density). A simple solution for integrated fields that
ensures the above requirements are satisfied is to express the field $F_{2}$ in normalised units -- i.e.
by dividing $F_{2}$ by its mean value. In this case, the variance of the normalised 3D density field can be
calculated from the variance of the normalised 2D column density field, as discussed below in Section~2.6.

\subsection{Approximations}

The power spectra of many fields of interest are power-law in form: 
$P_{3} \propto k^{-\alpha}$. If the power spectrum is steep (large $\alpha$) then the variance is sensitively
dependent on the power at low spatial frequencies. Because of the quantisation in equation~(\ref{eqno21cc}) and the
small amount of information available at low $k$, the above procedure may be inaccurate for large $\alpha$. 
Brunt \& Mac Low (2004) examined 3D and projected 2D standard deviations in the normalised density field 
($\rho/  \rho_{0}$) and normalised column density field ($ N / N_{0}$) obtained from
numerical simulations, with $\lambda = 128$. The (column) density power spectra were reasonably well fitted by
$P_{3}(k) \propto k^{-3}$, albeit with some curvature. They found that 
$\sigma_{N / N_{0}} / \sigma_{\rho / \rho_{0}} \approx 0.34 \pm 0.04$.
Evaluating equation~(\ref{eqno21cc}), using $\lambda = 128$ and $P_{3}(k) \propto k^{-3}$, to compute the ratio of 
standard deviations (i.e. $\sqrt{R}$), we predict $\sigma_{2} / \sigma_{3} = 0.39$, which is in acceptable
agreement with the Brunt \& Mac Low (2004) experimental result, given the spectral curvature.

With a power law form for the power spectrum, it is tempting to approximate equation~(\ref{eqno21cc}) with simple
integral-based expressions for $R$. If the spectral slope is $\alpha$ we could write:
\begin{equation}
R = \frac{2 \pi \int_{1}^{\lambda/2} \; dk \; k \;  k^{-\alpha}}{4 \pi \int_{1}^{\lambda/2} \; dk \; k^{2} \; k^{-\alpha}} ,
\label{eqno21a}
\end{equation}
which is easily solved for a specified $\alpha$. 
However, this is not a terribly good approximation as the integrals have been computed over circular and spherical regions of $k$-space
rather than the square and cubical regions which contain the fields. At low $\alpha$ this is not sufficiently accurate, and in general
we recommend that the direct summation method presented in equation~(\ref{eqno21cc}) be used. As a trivial example, for 
a cube of Gaussian noise ($\alpha = 0$), equation~(\ref{eqno21cc}) correctly predicts 
$R = (\lambda^{2} - 1)/(\lambda^{3} - 1)$, which
tends to $1/\lambda$ for large $\lambda$. In the same limit, equation~(\ref{eqno21a}) incorrectly predicts $R = 3/2\lambda$.
Equation~(\ref{eqno21a}) is useful however in gaining some intuitive understanding of how $R$ depends on the form
of the power spectrum: $R$ decreases if proportionally more power resides at large $k$ because the denominator is weighted by
$k^{2}$ and the numerator by $k$. The physical effect underlying this is that, if the variance is mostly at high $k$ (small spatial scales),
line-of-sight averaging suppresses more of the power than if the variance is mostly at low $k$ (large spatial scales).
Equation~(\ref{eqno21a}) is also useful for exploring scale-dependence effects on the calculated variances. Note that for 
$2 < \alpha \leq 3$, the 2D variance converges as $\lambda \longrightarrow \infty$ but the 3D variance diverges!

\subsection{Application to Observational Data}

For fields acquired through observations, it is necessary to account for the telescope's
beam response (point spread function) in the calculation of variances. The observational version of equation~(\ref{eqno21cc}) is:
\begin{equation}
R = \frac{{\displaystyle\sum_{k \neq 0}}^{2D,\lambda} P_{2}(k) \tilde{B}^{2}(k)}{{\displaystyle\sum_{k \neq 0}}^{3D,\lambda} P_{2}(k) \tilde{B}^{2}(k)}
\label{eqno22}
\end{equation}
where $\lambda$ is the number of pixels along each axis, $\tilde{B}^{2}(k) = \tilde{B}(k) \tilde{B}^{*}(k)$ is the square
of the Fourier space representation of the telescope beam pattern, and $P_{2}(k)$ is the power spectrum of
$F_{2}$ in the absence of beam-smoothing and instrumental noise. (Note that the observable quantity is
$P_{2}(k) \tilde{B}^{2}(k) + P_{2N}(k)$ where $P_{2N}(k)$ is the noise power spectrum.)
Equation~(\ref{eqno22}) includes accounting for the fact that the observed variance in the 2D field has been 
suppressed by the smoothing imposed by the telescope beam, and this must be taken into account.
In equation~(\ref{eqno22}), we have also applied the beam pattern to the denominator (representing the 3D variance). This
procedure may at first appear an odd choice, but it is motivated by the requirement that we limit our knowledge
of the 3D variance to the resolution provided by the data. In effect, equation~(\ref{eqno22}) is an attempt to
construct a 3D beam that samples the 3D density field and affords the same effective linear resolution as that 
provided by the telescope beam in projected 2D. The Fourier series $\tilde{F}_{3}$
extends to $k = \infty$, or at least to a (potentially very large) spatial
frequency, $k_{c}$, beyond which there are no variations in the field. 
In other words, there will be fluctuations in the projected 2D field (and therefore the 3D field) that we are 
not directly sensitive to, as they lie below our resolution limit. These, naturally, are additional sources of 
variance that we cannot measure. 

The full variance of the field $F_{3}$ should in principle be derived by 
removing (from the denominator) the effect of the beam pattern, and summing over the spatial frequency 
range $1 \leq k \leq k_{c}$.
In general, we will have little or no information about $k_{c}$, unless of course a 
distinct break in the power spectrum is observed. A possible, theoretically-motivated choice is that $k_{c}$ is the 
spatial frequency corresponding to the sonic scale (V\'{a}zquez-Semadeni et al 2003; Federrath et al 2009).  
We may try to account for this, by assuming a form for $P_{3}$ in the inaccessible range $k_{c} > k > \lambda/2$, 
by extending an observed power-law for example. This is quite 
dangerous, particularly for $\alpha \leq 3$ for which the summation diverges at large
spatial frequencies.
The case $\alpha = 3$ provides a convenient reference point, at which $R$ is underestimated by a factor
$\sim \ln{(\lambda/2)} / \ln{(k_{c})}$ -- a logarithmic divergence.
Without a direct constraint on $k_{c}$ it is more reasonable to restrict our knowledge of the 3D variance to the spatial dynamic 
range afforded by the observations, and utilize equation~(\ref{eqno22}) as stated, while recognizing that the 3D variance must be a 
lower limit to the true 3D variance. For numerical simulations, this is not an issue, as we have access to the full range of 
spatial frequencies at which structure is present. 

\subsection{Accounting for Non-Periodic Fields}

The Fourier series utilised above implicitly assume that the field is periodic.
While this is true of some numerical simulations, it is obviously not true of real physical fields.
If the field $F_{2}$ has a significant amplitude at the field boundaries, then elevated power
is produced in the 2D power spectrum at some spatial frequencies, typically along the
$k_{x} = 0$ and $k_{y} = 0$ axes, and this is not consistent with an isotropic form for
the 3D power spectrum. Firstly, we note that a signifcant amplitude of $F_{2}$ at the field
edges is almost certainly not consistent with the assumption of isotropy, as it implies that
$F_{2}$ has been extracted from a larger field with significant power on larger scales.
In this case, restriction of the field in the projected $x$ and $y$ directions is imposed
but there has been no corresponding restriction on the extent of the field in the $z$ direction.
Sensible definition of the field $F_{2}$ is therefore recommended, with, ideally, $F_{2}$
falling to near zero amplitude at the field boundaries.

It is possible to ameliorate the effect of edge discontinuities by edge-tapering the field
(e.g. Brunt \& Mac Low 2004). In this case we recommend that $\lambda$ is sufficiently
large that the tapering causes a negligible modification of the 2D variance. A good solution
is, in addition to tapering, to pad the field with zero values. If an observed field with
scale ratio $\lambda$ is zero-padded out to a larger size $\lambda_{p}$, then it is straightfoward
to show that the 3D variance should be calculated via:
\begin{equation}
\sigma_{3}^{2} = \frac{1}{\eta^{3}} (((\sigma_{2}^{2} + \langle F_{2} \rangle^{2})\eta^{2} - \langle F_{2} \rangle^{2} )/R_{p} + \langle F_{2} \rangle^{2}) - \langle F_{2} \rangle^{2} ,
\label{eqno22a}
\end{equation}
where $\eta = \lambda_{p}/\lambda$ and $R_{p}$ is the 2D-to-3D variance ratio calculated from the
power spectrum of the padded field. The quantities $\langle F_{2} \rangle$, $\sigma_{2}^{2}$, and $\sigma_{3}^{2}$ 
apply to the unpadded field.

Above we have assumed a square (cubical) box that contains the
field $F_{2}$ ($F_{3}$). In practice, we may wish to apply this method to fields that are
not exactly square, although this then obviously raises questions about the assumption of
isotropy. For a field of size $\lambda_{x} \times \lambda_{y}$, we recommend zero-padding
to produce a square field of size $\lambda_{px} \times \lambda_{py}$, and using equation~(\ref{eqno22a}) 
with $\eta = (\lambda_{px}\lambda_{py}/\lambda_{x}\lambda_{y})^{\frac{1}{2}}$. Note that this
assumes that the line-of-sight extent of $F_{3}$ is $\lambda_{z} = (\lambda_{x}\lambda_{y})^{\frac{1}{2}}$. 
Clearly, one should try to ensure that $\lambda_{x} \approx \lambda_{y}$ to respect the
assumption of isotropy.

\subsection{Summary of the Method}

To summarize the method: to estimate the 3D variance, $\sigma_{3}^{2}$, the following procedure should be
followed:
\begin{enumerate}
\renewcommand{\theenumi}{(\arabic{enumi})}
\item measure the variance, $\sigma_{2}^{2}$, of the normalised projected field $F_{2}$, \\
\item measure the power spectrum, $P_{2}(k)$, of $F_{2}$, and assume, through isotropy, that $P_{3}(k) \propto P_{2}(k)$,\\
\item using $P_{2}(k)$, compute the 2D-to-3D variance ratio, $R$, via equation~(\ref{eqno21cc}) or equation~(\ref{eqno22}),\\
\item compute $\sigma_{3}^{2}$, accounting for any zero-padding used to compute the power spectrum (equation~(\ref{eqno22a})). \\
\end{enumerate}

\subsection{Density Fields}

So far, other than isotropy, we have not imposed any particular structure to the field $F_{3}$
(e.g. by choosing a form for the power spectrum and PDF). It is well established that projected
column density power spectra of molecular clouds are power-law in form (e.g. Stutzki et al 1998; 
Bensch, Stutzki, \& Ossenkopf 2001), implying also that the 3D power spectra are also of power-law form. 
Typically, the 2D power spectrum is of the form $P_{2} \propto k^{-\alpha}$ where $\alpha \approx 3$. 
Density fields are always positive, and it is convenient to express both the density, $\rho$, and column density, $N$, 
in normalised units, by dividing by their respective mean values, $\rho_{0}$, and $N_{0}$, respectively. 
In this way, they will conform to the properties of the fields $F_{3}$ and $F_{2}$ discussed above (i.e. that
$F_{2}$ is the line of sight {\it average} of $F_{3}$). Otherwise, the column density
will be scaled by the physical length, $L$, as it is the integral of $\rho$ along the line-of-sight,
rather than the average. In most cases, our 2D observations will be of column density -- e.g. through
an optically thin spectral line, or through extinction mapping. 
The physical line-of-sight length, $L$, needed to convert column density to projected mean density, 
may be unknown. While $L$ may be inferred through the assumption of isotropy, if the distance is known, 
it is better to use the normalised density, $\rho / \rho_{0}$, and column density, $N / N_{0}$.

Using the observed variance, $\sigma_{N / N_{0}}^{2}$, in the normalised column density field, and the 
angular-averaged power spectrum, $P_{N/N_{0}}(k)$, we
can then use equation~(\ref{eqno22}) to obtain $R$. With appropriate treatment of any 
zero-padding (equation~(\ref{eqno22a}) with $\langle F_{2} \rangle = 1$, $\sigma_{2}^{2} = \sigma_{N / N_{0}}^{2}$,
and $\sigma_{3}^{2} = \sigma_{\rho / \rho_{0}}^{2}$),
the variance in normalised density, $\sigma_{\rho / \rho_{0}}^{2}$, can be calculated.

If a lognormal PDF in 3D is assumed, we can then derive $\sigma_{\ln{(\rho / \rho_{0})}}^{2}$ via:
\begin{equation}
\sigma_{\ln{(\rho / \rho_{0})}}^{2} = \ln{(1 + \sigma_{\rho / \rho_{0}}^{2})} .
\label{eqno25}
\end{equation}
Note that it is not necessary to assume a lognormal form for the column density PDF, although this indeed may 
be true (Ostriker, Stone, \& Gammie 2001; V\'{a}zquez-Semadeni \& Garc\'{i}a 2001).

Considering that $\sigma_{\rho / \rho_{0}}^{2}$ may be subject (for spectral slope $\alpha \approx 3$) to a 
logarithmic divergence because of unresolved 
density structure, as discussed in the previous section, equation~(\ref{eqno25}) has the fortunate property that the
computation of $\sigma_{\ln{(\rho / \rho_{0})}}^{2}$ suppresses this divergence to a log(log) divergence, so
that $\sigma_{\ln{(\rho / \rho_{0})}}^{2}$
can be quite well estimated even from observations at finite resolution, provided the spatial dynamic range is
sufficiently large to measure $\sigma_{N / N_{0}}^{2}$ accurately, and the field is of good linear
spatial resolution such that $\lambda/2$ approaches reasonable expectations for $k_{c}$. 
We recommend that estimates of the unresolved variance be made, using equation~(\ref{eqno21a}) as
a guide for suitable values of $k_{c}$ (e.g. the spatial frequency corresponding to the sonic scale, or even
$k_{c} \longrightarrow \infty$) to better assess the utility of the observational measurement (Brunt 2009). 
We note also that the above method for estimating $\sigma_{\rho / \rho_{0}}^{2}$ does not require the cloud 
distance to be known. However, observations at fixed angular resolution applied to clouds at different distances 
will probe different physical scales, so the method is not entirely distance-independent.

The above prescription can, in principle, be applied to other positive-definite fields, such as temperature.
Estimation of mean projected temperature fields is usually
done by quite a different procedure than that used for column density fields, however. Typically, we are
able to derive a line-of-sight average temperature for each pixel, either through flux ratios at far infrared
wavelengths, or by excitation analysis of millimetre-wave spectral lines, for example. These measurements are
often not straightforward {\it spatial} averages, as assumed by our method, but may be (e.g.) density-weighted averages
instead. Some caution must be applied to the treatment of such fields.

\subsection{Velocity Fields}

Application of the method to projected mean velocity fields is also possible. Obviously, since
a velocity field is not positive-definite, we do not normalize the field as was done for the density field.
In principle, we do have access to the mean velocity field, which can be obtained through an optically 
thin spectral line. However, such a field is density-weighted, rather than a direct spatial average
as assumed by equation~(\ref{eqno2}), and significant problems in estimating the power spectrum (Brunt \& Mac Low 2004), 
and presumably therefore the 2D variance, can arise for supersonic turbulence. There will be lines of sight
with insufficient signal-to-noise to calculate a mean velocity. Therefore the variance must be calculated for the
detected lines of sight, with the assumption that these reliably represent the entire field. Note that the
problem in estimating the mean of the field (as required in the case of column density / density analysis) 
does not apply here. We also have access to only
one component of the velocity field (the line-of-sight component $v_{z}$) and we must assume not only
isotropy in this, but also apply the isotropic assumption to the entirely inaccessible $v_{x}$ and $v_{y}$
components. With these provisos, the method would proceed as outlined at the end of Section 2.1.

\begin{figure}
\includegraphics[width=74mm]{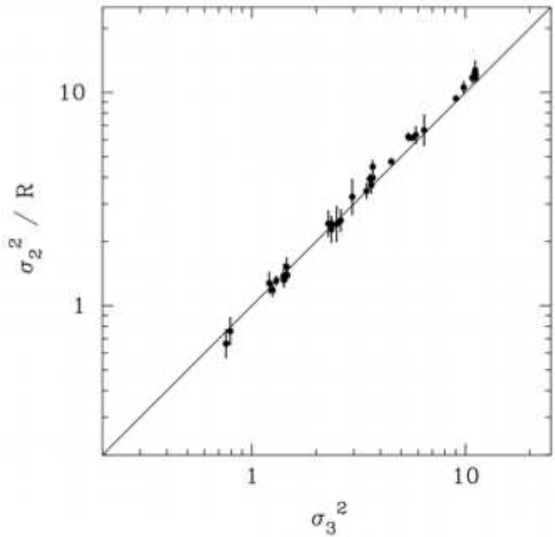}
%  \vspace*{174pt}
  \caption{Plot of the estimated 3D density variance, $\sigma_{2}^{2} / R$, versus
the actual 3D density variance, $\sigma_{3}^{2}$, obtained from numerical simulations
of hydrodynamic turbulence. Each plotted point represents the mean value of
$\sigma_{2}^{2} / R$ obtained from analysis of column density fields projected along the
three cardinal directions, while the error bars represent the standard deviation in the
values of $\sigma_{2}^{2} / R$ for the three directions.}
\label{fig:1}
\end{figure}

There is a rather more straightforward method of estimating the line-of-sight
velocity dispersion, $\sigma_{v_{z}}$, which is
simply to measure the dispersion of the mean line profile obtained through imaging observations
of an optically-thin spectral line. (This is also a density-weighted measure, however.) If we have an estimate 
of the cloud temperature, and an estimate of the mean molecular weight, we can then derive the Mach number to
construct the right-hand-side of equation~(\ref{eqno0}).

\section{Application to Numerical Simulations}

As an initial test of the method, we now apply it to numerical simulations
of hydrodynamic and magnetohydrodynamic turbulence. 
Given that the above theory is derived assuming isotropic fields, we 
investigate the effect of anisotropy using magnetohydrodynamic simulations.
In all the studies below, we use normalised column densities ($N / N_{0}$) 
and normalised densities ($\rho / \rho_{0}$) so that $\sigma_{2}^{2} = \sigma_{N / N_{0}}^{2}$
and $\sigma_{3}^{2} = \sigma_{\rho / \rho_{0}}^{2}$.

\begin{figure}
\includegraphics[width=74mm]{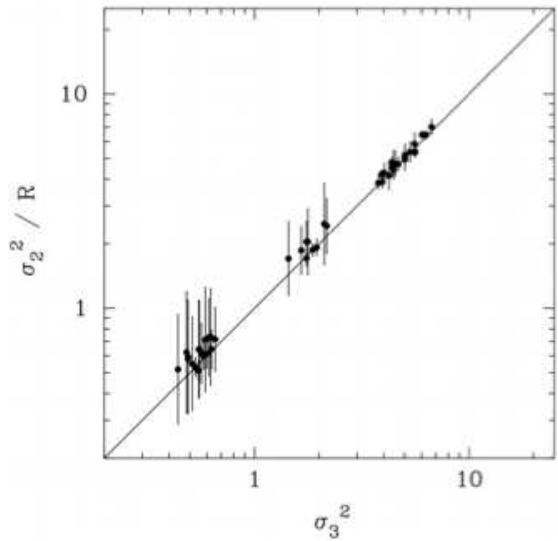}
% \vspace*{174pt}
  \caption{Plot of the estimated 3D density variance, $\sigma_{2}^{2} / R$, versus
the actual 3D density variance, $\sigma_{3}^{2}$, obtained from numerical simulations
of magnetohydrodynamic turbulence. Each plotted point represents the mean value of
$\sigma_{2}^{2} / R$ obtained from analysis of column density fields projected along the
three cardinal directions, while the error bars represent the standard deviation in the
values of $\sigma_{2}^{2} / R$ for the three directions. The solid line is the line of equality.}
\label{fig:2}
\end{figure}

\begin{figure}
\includegraphics[width=74mm]{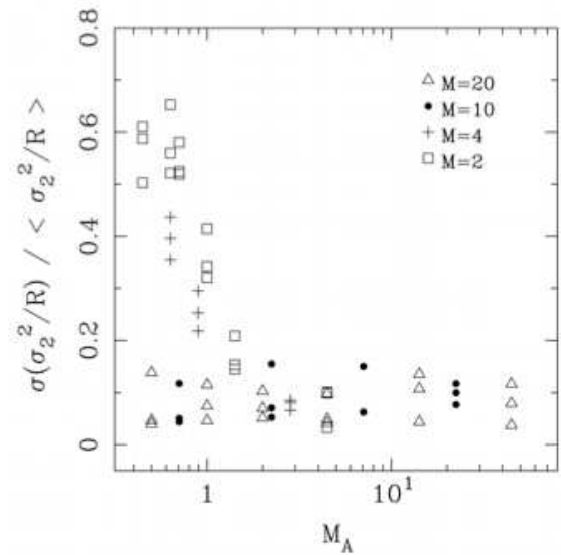}
%  \vspace*{174pt}
  \caption{Plot of the standard deviation in $\sigma_{2}^{2} / R$ expressed as
a fraction of the mean $\sigma_{2}^{2} / R$, versus the Alfv\'{e}nic Mach number, $M_{A}$,
for the MHD simulations. }
\label{fig:3}
\end{figure}

\subsection{Hydrodynamic Turbulence}

First, we analyse hydrodynamic simulations at a range of rms Mach numbers: 1.25, 2.5, 3.5, 5, 7, 10, and 20.
Multiple snapshots are used at each Mach number setting, separated by at least a crossing time.
The hydrodynamic simulations were run with the \textsc{phantom} Smoothed Particle Hydrodynamics code, with turbulence driven
artificially over ~5 crossing times using large-scale solenoidal Fourier driving (Price \& Federrath 2009;
Federrath et al 2009; Brunt, Heyer, \& Mac Low 2009; Brunt 2003).
For the analysis in this paper we performed a suite of low resolution $128^3$ particle calculations,
interpolated to $256^3$ grids. Since the key idea here is the reconstruction of the density variance, we find
that even low resolution calculations are sufficient as demonstrators of the method.
For each simulation, we take the 3D density field, of mean unity, and create three normalised
projected fields by averaging along each spatial axis. We compute $\sigma_{3}^{2}$ in 3D and a value
of $\sigma_{2}^{2}$ for each projected 2D field.

\begin{figure*}
\includegraphics[width=170mm]{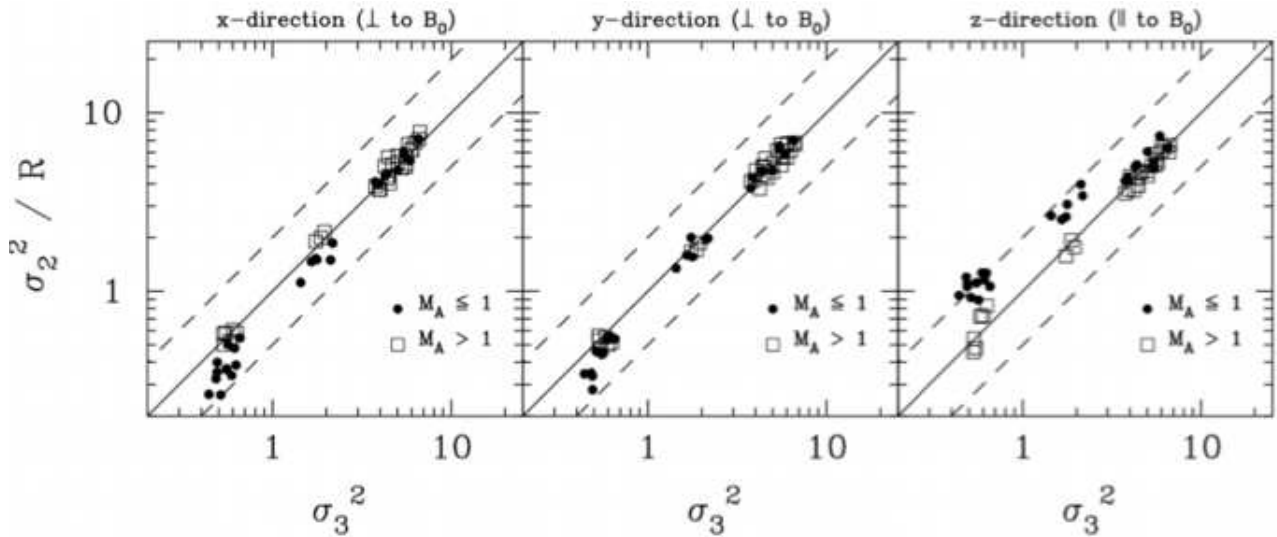}
%  \vspace*{174pt}
  \caption{ Relations between $\sigma_{2}^{2} / R$ and the true $\sigma^{2}_{3}$ for each of the
three axes. Runs with $M_{A} \leq 1$ are shown as dots; runs with $M_{A} > 1$ are shown as open
squares. The solid line is the line of equality and the dotted lines mark underestimation
and overestimation of $\sigma^{2}_{3}$ by a factor of 2.
 }
\label{fig:4}
\end{figure*}

The power spectrum for each column density field is then measured (specifically,
we compute the average power in bins of unit width in $k$.) Following this, equation~(\ref{eqno21cc})
is used to compute a value of $R$ for each projected field. To do this, we interpolate the
binned column density power spectrum to the appropriate power for each $k$ resulting from
the nested sums. An extrapolation is required (using a power law fit to the
power spectrum in the high $k$ region) for $\sqrt{2} \lambda/2 < k \leq \sqrt{3} \lambda/2$ in the
denominator. In the above procedure, we emphasize that only information present in the column density
field is used in the calculation of $R$. Values of $R$ computed for these fields vary
between about 0.03 and 0.15. Note that in general, $R$ will depend on the form of the power spectrum
and the scale ratio of the field.

In figure~\ref{fig:1} we plot the estimated 3D density variance, $\sigma_{2}^{2} / R$, versus
the actual 3D density variance, $\sigma_{3}^{2}$. For each snapshot, we represent the
measurements as a mean (plotted point) and standard deviation (error bar) obtained from the
three different projections. The method can predict the 3D density variance to about 10\% accuracy.

\subsection{Magnetohydrodynamic Turbulence}

\begin{figure}
\includegraphics[width=74mm]{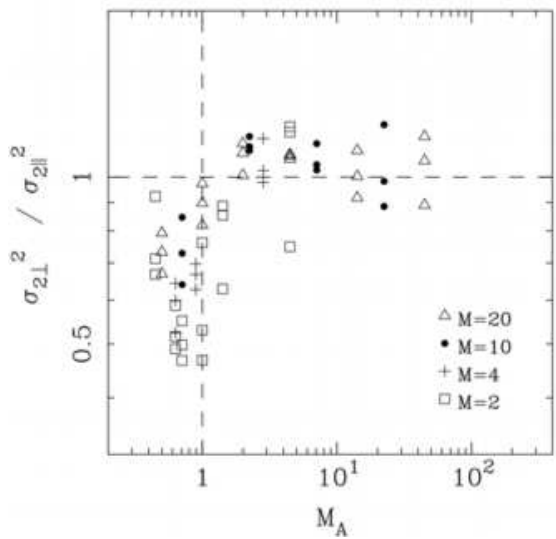}
%  \vspace*{174pt}
  \caption{Plot of the anisotropy indicator, $\sigma^{2}_{2\perp}/\sigma^{2}_{2||}$, versus the Alfv\'{e}nic Mach number, $M_{A}$,
for the MHD simulations. For reference, the horizontal dashed line marks the expected value of
$\sigma^{2}_{2\perp}/\sigma^{2}_{2||}$ for an isotropic field; the vertical dashed line marks $M_{A} = 1$, below
which anisotropy is notable.}
\label{fig:5}
\end{figure}

We now analyse magnetohydrodynamic (MHD) simulations of turbulence. 
The MHD simulations were run with the grid-based code \textsc{flash} (Fryxell et
al. 2000). 
We used a new approximate Riemann solver for ideal-MHD (Bouchut, Klingenberg, \& Waagan 2007, 2009), 
which preserves positive states in highly supersonic MHD turbulence. This solver was recently 
developed for \textsc{flash} by Waagan (2009). The corresponding scheme for
preserving positive states in hydrodynamical studies has been
successfully tested and applied in Klingenberg, Schmidt \& Waagan (2007).
We used the same solenoidal forcing scheme as used in Federrath et al.
(2009) and Price \& Federrath (2009). All MHD models were evolved on a
fixed grid with $256^3$ grid zones.

\begin{figure*}
\includegraphics[width=170mm]{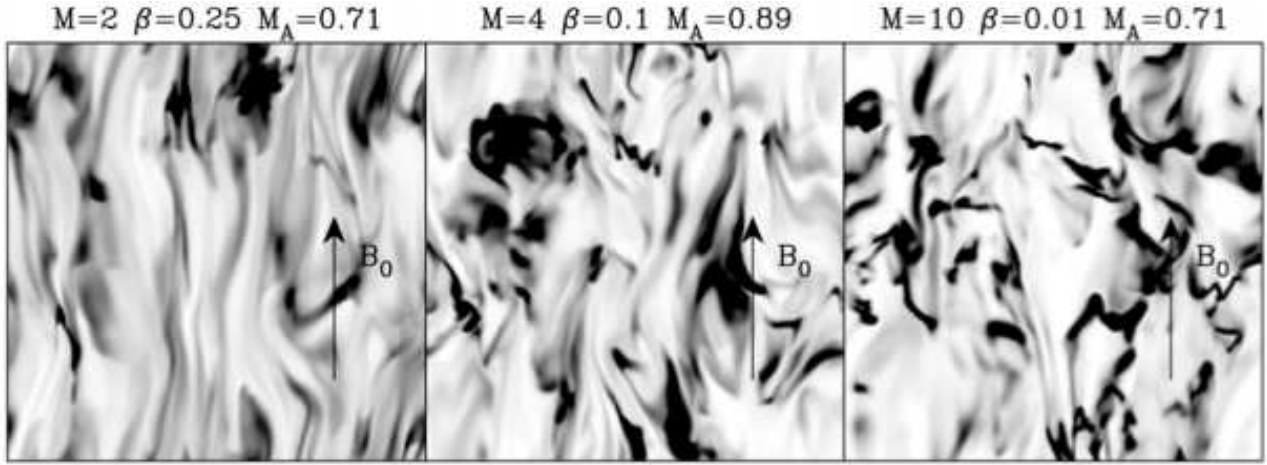}
%  \vspace*{174pt}
  \caption{ Representative 2D slices through a sample of density fields created by sub-Alfv\'{e}nic
turbulence. The mean magnetic field direction is illustrated by an arrow, labelled $B_{0}$. The
values of the sonic Mach number, $M$, the plasma beta, $\beta$, and the Alfv\'{e}nic Mach
number, $M_{A}$ are given above each panel.
}
\label{fig:6}
\end{figure*}

\begin{figure*}
\includegraphics[width=170mm]{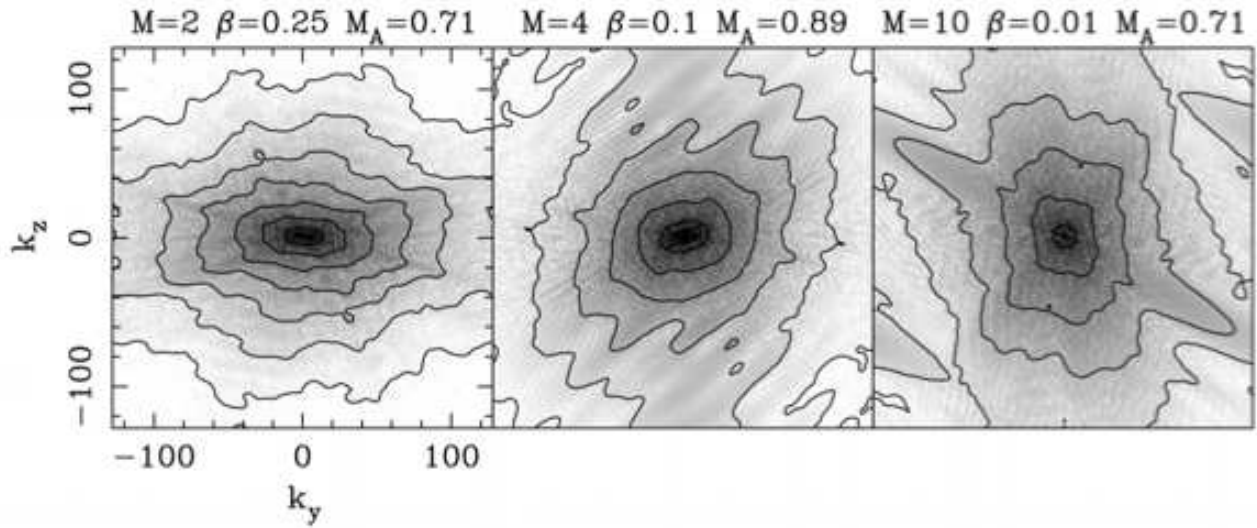}
%%  \vspace*{174pt}
  \caption{ Power spectra of the 2D density field slices from Figure~\ref{fig:6}. Greyscales are
represented in logarithmic units, with contours, each separated by a decade in power, overlayed; the
contours been smoothed for clarity.
}
\label{fig:7}
\end{figure*}

In addition to varying the Mach number, in the MHD simulations we now also vary the ratio 
of thermal to magnetic pressure, $\beta = P/(\frac{B_{0}^{2}}{2\mu_{0}})$. For large magnetic field strengths (lower $\beta$)
significant anisotropy can arise in the density and velocity fields (Mac Low 1999; Vestuto, Ostriker, \& Stone 2003;
Heyer et al. 2008), and thus the assumptions of our method could break down. Using the MHD density fields,
we repeat the same procedure described above to calculate $R$, and hence the estimated 3D variance,
$\sigma^{2}_{2} / R$; the comparison of estimated to true 3D variance is shown in Figure~\ref{fig:2}.
We find that the method is still accurate in predicting the mean (plotted points) but the variation
between the different directions (error bars) is now significantly larger for some parameterizations. 
Investigation of this shows that significant variations in $\sigma^{2}_{2}/R$ between the different
projection axes are seen if the Alfv\'{e}nic Mach number, $M_{A} = M\sqrt{(\beta/2)}$, is
smaller than $\sim$~unity. 
To show this, in Figure~\ref{fig:3} we plot the fractional error in the estimated 3D density variance
(i.e the standard deviation in $\sigma_{2}^{2} / R$ divided by the mean $\sigma_{2}^{2} / R$ obtained from the
three different projection directions) versus $M_{A}$. The fractional error in the estimated 3D density
variance is around 10\% for $M_{A} > 1$, while it increases significantly for $M_{A} < 1$.

As there is now notable variation in $\sigma_{2}^{2} / R$ between the different directions, it is 
worth plotting the individual $\sigma_{2}^{2} / R$ versus $\sigma^{2}_{3}$ relations obtained 
from each axis. These relations are shown in Figure~\ref{fig:4}, where we have distinguished 
the plotted points with different symbols for $M_{A} \leq 1$ and $M_{A} > 1$. The 3D variance
in sub-Alfv\'{e}nic runs is overestimated (underestimated) from 2D fields produced by averaging over
an axis parallel (perpendicular) to the mean magnetic field direction, by as much as a factor
of $\sim$~2.

\begin{figure*}
\includegraphics[width=170mm]{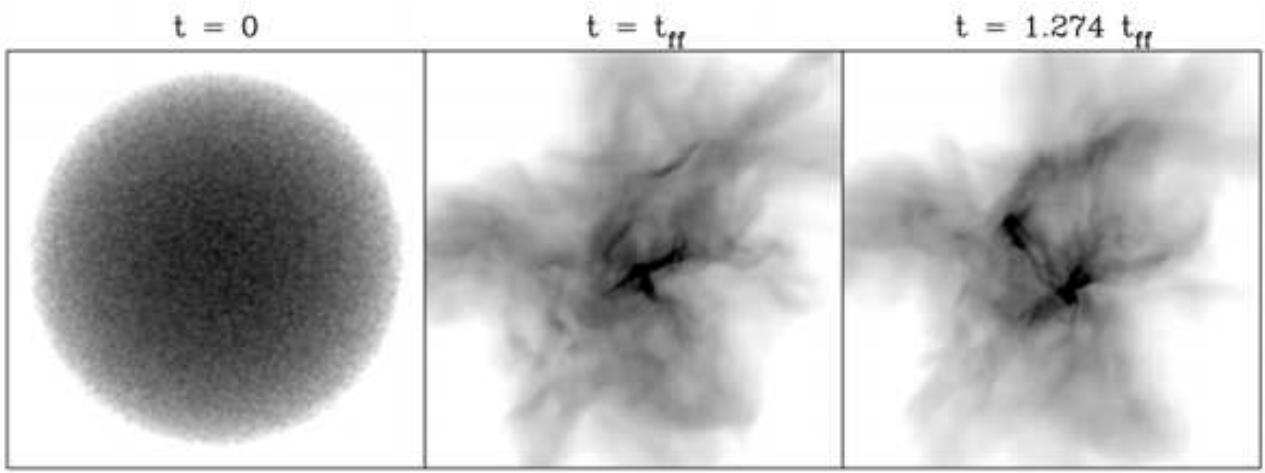}
%%  \vspace*{174pt}
  \caption{ Column density plots (logarithmic greyscale) of the non-periodic SPH simulation.
}
\label{fig:8}
\end{figure*}

The variations in $\sigma_{2}^{2} / R$ are caused by anisotropic structure in the density field that is
produced when $M_{A} < 1$. As a simple measure of this anisotropy, we define an ``anisotropy indicator''
using the 2D fields that are produced by averaging over the three cardinal axes.  
The anisotropy indicator is defined
as $\sigma^{2}_{2\perp}/\sigma^{2}_{2||} = (\sigma^{2}_{2x} + \sigma^{2}_{2y})/2\sigma^{2}_{2z}$,
where $\sigma^{2}_{2x}$, $\sigma^{2}_{2y}$, and $\sigma^{2}_{2z}$ are the variances in the
2D fields produced by averaging over the $x$, $y$, and $z$ axes respectively. The
measured values of $\sigma^{2}_{2\perp}/\sigma^{2}_{2||}$ are plotted in Figure~\ref{fig:5}
versus $M_{A}$. For $M_{A} > 1$, $\sigma^{2}_{2\perp}/\sigma^{2}_{2||}$ is $\sim$~unity,
indicating roughly equal variances in each of the projected 2D fields. For
$M_{A} < 1$ the variance in the 2D field averaged over the $z$ direction
(parallel to the mean $B$ field) is larger, by as much as a factor of 2, than that of 
the 2D fields produced by averaging over the $x$ or $y$ directions 
(perpendicular to the mean $B$ field).

Interestingly, the degree of anisotropy appears to be related to the sonic Mach number, $M$,
as is evident in both Figure~\ref{fig:3} and Figure~\ref{fig:5}. Indeed, for $M \geq 10$, the
estimation of the 3D variance is not noticeably worse for sub-Alfv\'{e}nic turbulence
than for super-Alfv\'{e}nic turbulence, and the largest deviations from isotropic
behaviour are found in the $M = 2$ runs. A direct visualization of the anisotropies
present in the density fields is given in Figure~\ref{fig:6}, which shows representative
2D slices through a sample of density fields produced by sub-Alfv\'{e}nic turbulence. At low
sonic Mach number ($M = 2$) the density field is characterized by many long filamentary
structures oriented parallel to the mean magnetic field direction. As the sonic Mach number
is increased, this directional order in the density field is systematically reduced, until
at $M = 10$ it is barely noticeable. The power spectra of these 2D density field slices 
are shown in Figure~\ref{fig:7}. These confirm our visual assessment of the density field slices. The
high frequency spectral power along the magnetic field axis is notably lower in the low sonic 
Mach number calculations.

The tests on the MHD simulations provide a valuable baseline for establishing the physical
regimes in which our method can be applied. In most circumstances encountered in molecular
clouds with sizes of a few parsecs, sonic Mach numbers are large enough to enforce sufficient
isotropy that the method can work with around 10\% accuracy, even for sub-Alfv\'{e}nic
turbulence. However, these are rather idealized conditions, as the models do not include
other relevant physics such as self-gravity.
Gravitational collapse in the presence of a magnetic field is likely to be oriented along
the magnetic field lines if the field is sufficiently strong. Note that the combination
of magnetic fields and gravity, in this instance, can induce anisotropic structure in
the orthogonal direction to that seen in the sub-Alfv\'{e}nic simulations described above.

\begin{figure}
\includegraphics[width=74mm]{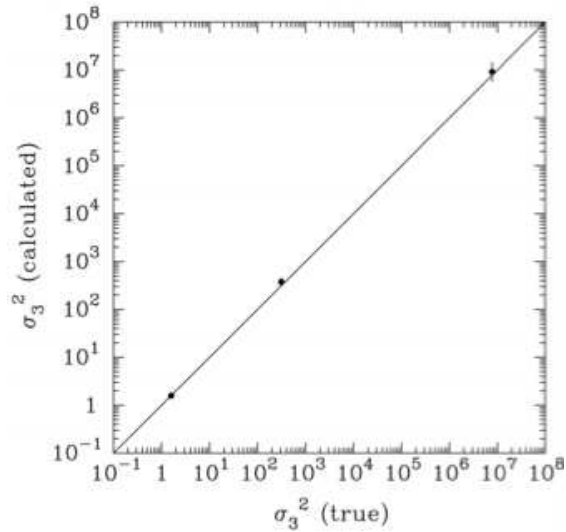}
%%  \vspace*{174pt}
  \caption{ Comparison of calculated to true 3D variances for the non-periodic SPH density fields.
}
\label{fig:9}
\end{figure}

\subsection{Non-Periodic Fields}

In the foregoing analysis, we have used periodic fields for which there are no edge discontinuities
and consequently no zero-padding is required. A more physically realistic scenario is provided
by SPH simulations that do not employ periodic boundary conditions. For our analysis, we take
density fields produced by the hydrodynamic simulations of Price \& Bate (2009). Column 
density fields extracted at three different times (at $t = 0$; after one free fall time;
and at the end of the calculation where $t$~=~1.274~$t_{ff}$) are shown in Figure~\ref{fig:8}. (The
column density fields are represented as logarithmic greyscales as the variances are extremely large.)
We zero-pad these fields to $\lambda_{p} = 512$ from the original $\lambda = 256$, then compute the power
spectra and from this derive $R_{p}$. We then calculate the 3D variance by use of equation~(\ref{eqno22a}).

In Figure~\ref{fig:9} we compare the 3D variances derived from our method with the actual 3D variances
measured directly from the density fields. We find that the method is comparably accurate for these 
non-periodic fields, albeit with a rather large variation ($\sim$~40\%) between different projections
in the latest snapshot. In this field, the variance is very large indeed ($7.7\times10^{6}$) and is dominated
by a small number of very large density values caused by localised gravitational collapse. The variances in the initial
density field (a spherical blob), in the $t = t_{ff}$ density field, and in the $t = 1.274~t_{ff}$ density field 
are recovered by the mean calculated variances to within 0.3\%, 20\% and 20\% accuracy respectively. 
To evaluate the success of these measurements, we calculate an anisotropy indicator as was done in
Section~3.2, by taking the ratio of the variance in one projected field to the mean of the variances in the
other two projected fields. Since the simulations contain no magnetic fields, there is no preferred axis. Therefore
we combine the three 2D variance measurements in a way that minimises the anisotropy indicator (lower values
represent greater degrees of anisotropy). For the three
snapshots, in increasing-time-order, we find anisotropy indicators of 1.00, 0.78, and 0.54. For the two
evolved snapshots, the density fields are comparably anisotropic to the low $M_{A}$, low $M$ density fields of 
the previous section, but with the anisotropy now arising from the initial velocity perturbations rather than from
the effect of a magnetic field. The accuracy of the method for non-periodic boundary conditions
is therefore comparable to the periodic case when compared at the same level of intrinsic anisotropy. 

The extremely high variances seen in the evolved density fields are a result of small-scale collapse in this
gravitationally-bound cloud, and far exceed variances expected by equation~(\ref{eqno0}) with $b \approx 0.5$. 
It will be interesting in future to see how equation~(\ref{eqno0}) is modified in strongly-self-gravitating clouds,
and whether observations can quantify this.

\section{Conclusions}

We have introduced and tested a simple method for measuring the 3D density variance in molecular
clouds, using only information present in the projected column density field. The method relies on
the assumption of isotropy and uses the measured column density power spectrum in conjunction with Parseval's 
theorem to calculate a correction factor, $R$, that scales the observed normalised column density variance to
the intrinsic three-dimensional normalised density variance. The method is sufficiently general to be applied to
any isotropic field.

For density fields produced in supersonic hydrodynamic and magnetohydrodynamic turbulence, the method is 
accurate to about 10\% provided that the assumption of isotropy is valid. 
For turbulent density fields, in practice this requires that the turbulence motions are super-Alfv\'{e}nic ($M_{A} < 1$), 
though even in the sub-Alfv\'{e}nic regime we are able to recover the 3D variance for high sonic Mach number ($M \gtrsim 10$).

\section*{Acknowledgments}
This work was supported by STFC Grant ST/F003277/1 and Marie Curie Re-Integration Grant MIRG-46555.
CB is supported by an RCUK fellowship at the University of Exeter, UK. 
CF is grateful for financial support by the International Max Planck
Research School for Astronomy and Cosmic Physics (IMPRS-A) and the
Heidelberg Graduate School of Fundamental Physics (HGSFP), which is
funded by the Excellence Initiative of the German Research Foundation
(DFG GSC 129/1). The \textsc{FLASH} MHD simulations were run at the
Leibniz-Rechenzentrum in Munich (grant pr32lo). The software used in
this work was in part developed by the DOE-supported ASC / Alliance
Center for Astrophysical Thermonuclear Flashes at the University of
Chicago. 
We thank the anonymous referee for useful and interesting suggestions.

%\appendix

\label{lastpage}


\begin{thebibliography}{99}
\bibitem[\protect\citeauthoryear{Bensch, Stutzki, \& Ossenkopf}{2001}]{bso01} Bensch, F., Stutzki, J., \& Ossenkopf, V., 2001, A{\&}A, 366, 636
\bibitem[{{Bouchut} {et~al.}(2007){Bouchut}, {Klingenberg}, \& {Waagan}}]{BouchutKlingenbergWaagan2007} {Bouchut}, F., {Klingenberg}, C., \& {Waagan}, K. 2007, Numerische Mathematik, 108, 7
\bibitem[{{Bouchut} {et~al.}(2009){Bouchut}, {Klingenberg}, \& {Waagan}}]{BouchutKlingenbergWaagan2009} {Bouchut}, F., {Klingenberg}, C., \& {Waagan}, K. 2009, Numerische Mathematik, accepted
\bibitem[\protect\citeauthoryear{Brunt}{2009}]{b09} Brunt, C. M., 2009, submitted to A{\&}A
\bibitem[\protect\citeauthoryear{Brunt}{2003}]{b03} Brunt, C. M., 2003, ApJ, 583, 280
\bibitem[\protect\citeauthoryear{Brunt, Heyer, \& Mac Low}{2009}]{bhml09} Brunt, C. M., Heyer, M. H., \& Mac Low, 2009, A{\&}A, 504, 883
\bibitem[\protect\citeauthoryear{Brunt \& Mac Low}{2004}]{bml04} Brunt, C. M., \& Mac Low, 2004, ApJ, 604, 196 
\bibitem[\protect\citeauthoryear{Elmegreen}{2008}]{e08}	Elmegreen, B. G., 2008, ApJ, 672, 1006
\bibitem[\protect\citeauthoryear{Federrath, Klessen, \& Schmidt}{2008}]{f08} Federrath, C., Klessen, R. S., \& Schmidt, W., 2008, ApJ, 688, 79
\bibitem[\protect\citeauthoryear{Federrath et al}{2009}]{fdksm09} Federrath, C., Duval, J., Klessen, R., Schmidt, W., \& Mac Low, M.-M., 2009, arXiv, 0905.1060
\bibitem[\protect\citeauthoryear{Fryxell et al}{2000}]{f00} Fryxell, B., Olson, K., Ricker, P., Timmes, F. X., Zingale, M., Lamb, D. Q., MacNiece, P., Rosner, R., Truran, J. W., \& Tufo, H., 2000, ApJS, 131, 273
\bibitem[\protect\citeauthoryear{Goodman, Pineda, \& Schnee}{2009}]{gps09} Goodman, A. A., Pineda, J. E., \& Schnee, S. L., 2009, ApJ, 692, 91
\bibitem[\protect\citeauthoryear{Hartmann}{2002}]{hart02} Hartmann, L., 2002, ApJ, 578, 914
\bibitem[\protect\citeauthoryear{Hennebelle \& Chabrier}{2008}]{hc08} Hennebelle, P., \& Chabrier, G., 2008, ApJ, 684, 395
\bibitem[\protect\citeauthoryear{Hennebelle \& Chabrier}{2009}]{hc09} Hennebelle, P., \& Chabrier, G., 2009, ApJ, 702, 1428
\bibitem[\protect\citeauthoryear{Heyer et al}{2008}]{hgob08} Heyer, M. H., Gong, H., Ostriker, E., \& Brunt, C., 2008, ApJ, 680, 420
%\bibitem[\protect\citeauthoryear{Hildebrand}{1983}]{hil83} Hildebrand, R. H., 1983, QJRAS, 24, 267
\bibitem[\protect\citeauthoryear{Klingberg, Schmidt, \& Waagan}{2007}]{csw07} Klingberg, C., Schmidt, W., \& Waagan, K., 2007, J. Comp. Phys., 227, 12
\bibitem[\protect\citeauthoryear{Krumholz \& McKee}{2005}]{km05} Krumholz, M. R., \& McKee, C. F., 2005, ApJ, 630, 250
\bibitem[\protect\citeauthoryear{Lazarian \& Pogosyan}{2000}]{lp00} Lazarian, A., \& Pogosyan, D., 2000, ApJ, 537, 720
\bibitem[\protect\citeauthoryear{Lucy}{1974}]{l74} Lucy, L. B., 1974, AJ, 79, 745
\bibitem[\protect\citeauthoryear{Mac Low}{1999}]{ml99} Mac Low, M.-M., 1999, ApJ, 524, 169
\bibitem[\protect\citeauthoryear{Miville-Deschenes \& Martin}{2007}]{mvdm07} Miville-Deschenes, M.-A., \& Martin, P. G., 2007, A{\&}A, 469, 189
\bibitem[\protect\citeauthoryear{Ostriker, Stone, \& Gammie}{2001}]{osg01} Ostriker, E. C., Stone. J. M., \& Gammie, C. F., 2001, ApJ, 546, 980
\bibitem[\protect\citeauthoryear{Padoan, Nordlund, \& Jones}{1997}]{pnj97} Padoan, P., Nordlund, \AA., \& Jones, B. J. T., 1997, MNRAS, 288, 145
\bibitem[\protect\citeauthoryear{Padoan \& Nordlund}{2002}]{pn02} Padoan, P., \& Nordlund, \AA., 2002, ApJ, 576, 870
\bibitem[\protect\citeauthoryear{Padoan \& Nordlund}{2009}]{pn09} Padoan, P., \& Nordlund, \AA., 2009, arXiv, 0907.0248
\bibitem[\protect\citeauthoryear{Passot \& V\'{a}zquez-Semadeni}{1998}]{pv98} Passot, T., \& V\'{a}zquez-Semadeni, E., 1998, PhRvE, 58, 4501
\bibitem[\protect\citeauthoryear{Price \& Bate}{2009}]{pb09} Price, D. J., \& Bate, M. R., 2009, MNRAS, 398, 33
\bibitem[\protect\citeauthoryear{Price \& Federrath}{2009}]{pf09} Price, D. J., \& Federrath, C., 2009, submitted to MNRAS
\bibitem[\protect\citeauthoryear{Reblinsky}{2000}]{r2000} Reblinsky, K., 2000, A{\&}A, 364, 377
\bibitem[\protect\citeauthoryear{Stutzki et al}{1998}]{s98} Stutzki, J., Bensch, F., Heithausen, A., Ossenkopf, V., \& Zeilinsky, M., 1998, A{\&}A, 336, 697
\bibitem[\protect\citeauthoryear{V\'{a}zquez-Semadeni}{1994}]{v94} V\'{a}zquez-Semadeni, E., 1994, ApJ, 423, 681
\bibitem[\protect\citeauthoryear{V\'{a}zquez-Semadeni, Ballesteros-Paredes, \& Klessen}{2003}]{vbpk03} V\'{a}zquez-Semadeni, E., Ballesteros-Paredes, J., \& Klessen, R. S., 2003, ApJ, 585, L131
\bibitem[\protect\citeauthoryear{V\'{a}zquez-Semadeni \& Garc\'{i}a}{2001}]{vg01} V\'{a}zquez-Semadeni, E., \& Garc\'{i}a, N., 2001, ApJ, 557, 727
\bibitem[\protect\citeauthoryear{Vestuto, Ostriker, \& Stone}{2003}]{vos03} Vestuto, J. G., Ostriker, E. C., \& Stone, J. M., 2003, ApJ, 590, 858
\bibitem[{{Waagan}(2009)}]{Waagan2009} {Waagan}, K. 2009, Journal of Computational Physics, 228, 860
\end{thebibliography}
\end{document}